\documentclass[twocolumn,
prl,
showpacs,floats,preprintnumbers,amsmath,amssymb]{revtex4}
\usepackage{epsfig}
\usepackage{amsmath}

\renewcommand{\[}{\left [}
\renewcommand{\]}{\right ]}
\def\fslash#1{#1 \!\!\! \slash}
\def\beq{\begin{equation}}
\def\eeq{\end{equation}}
\def\pa{\partial}

\def\bea{\arraycolsep .1em \begin{eqnarray}}
\def\eea{\end{eqnarray}}

\def\vp{{\bf p}}

\def\vk{{\bf k}}

\def\Tr{{\rm Tr}}

\let\si=\sigma

\let\om=\omega

\let\no=\nonumber

\def\et{{\em et al.}}
\def\eq#1{Eq.(\ref{#1})}
\def\refr#1{\cite{#1}}

\def\s0#1#2{\mbox{\small{$ \frac{#1}{#2} $}}}
\def\0#1#2{\frac{#1}{#2}}

\def\anp#1#2#3{Adv.\ Nucl.\ Phys. \ {\bf #1}, #2 (#3)}
\def\plb#1#2#3{Phys. Lett. {\bf B #1}, #2 (#3)}
\def\npa#1#2#3{Nucl. Phys. {\bf A #1}, #2 (#3)}

\def\prc#1#2#3{Phys. Rev.  {\bf C #1}, #2 (#3)}
\def\prd#1#2#3{Phys. Rev. {\bf D #1}, #2 (#3)}
\def\prl#1#2#3{Phys. Rev. Lett. {\bf #1}, #2 (#3)}
\def\ann#1#2#3{Ann. Phys. (N.Y.) {\bf #1}, #2 (#3)}
\def\anp#1#2#3{Adv. Nucl. Phys. {\bf #1}, #2 (#3)}

\def\prep#1#2#3{Phys.\ Rep.\ {\bf #1}, #2 (#3)}

\def\ijmpe#1#2#3{Int.\ J.\ Mod.\ Phys.\ {\bf E #1}, #2 (#3)}
\def\jhep#1#2#3{J.\ High Energy Phys.\ {\bf #1}, #2 (#3)}
\def\rmp#1#2#3{Rev.\ Mod.\ Phys.\ {\bf #1}, #2 (#3)}
\def\zpa#1#2#3{Z.\ Phys.\ {\bf A #1}, #2 (#3)}

\def\ibid#1#2#3{{\it ibid.}, {\bf #1}, #2 (#3)}
\begin{document}

\title{
Novel effects of electromagnetic interaction on the correlation of nucleons
in nuclear matter}
\author{Ji-sheng Chen\footnote{chenjs@iopp.ccnu.edu.cn},
Jia-rong Li, 
and Meng Jin
}
\affiliation{
Institute of Particle Physics and Physics Department, Hua-Zhong
Normal University, Wuhan 430079, People's Republic of China}
\begin{abstract}
The electromagnetic(EM) interactions between charged protons on the correlations of nucleons
    are discussed by introducing the Anderson-Higgs mechanism of broken $U(1)$ EM symmetry
    into the relativistic nuclear theory with a parametric photon mass.
The non-saturating Coulomb force contribution is emphasized on the equation of state
    of nuclear matter with charge symmetry breaking(CSB) at finite temperature
    and the breached $^1S_0$ pairing correlations of  proton-proton and neutron-neutron.
The universal properties given by an order parameter field with a non-zero
    vacuum expectation value (VEV) nearby phase transition are explored within the mean field
    theory(MFT) level.
This mechanism can be extended to the charged or charge neutralized strongly
    coupling multi-components system for the discussion of binding or pairing issues.

\end{abstract}
\pacs{21.30.Fe, 21.10.Sf, 11.30.Cp}
\maketitle
Understanding the properties of nuclear matter under both normal and extreme conditions
    is of great importance in relativistic heavy ion
    collisions and explaining the appearance of compact objects such as the
    neutron stars and neutron-rich matter or nuclei.
The determination of the properties of nuclear matter as functions of
    density/temperature, the ratio of protons to neutrons, and
    the pairing correlations-superfluidity or superconductivity
    is a fundamental problem in contemporary physics\refr{danielewicz2002,dean2003}.
The discussion about the property of nuclear ground
    state-binding energy and pairing correlations at low temperature is substantial.

The in-medium behavior associated with the many-body characteristic is the key, while a non-perturbative approach is crucial.
The theoretical difficulty of making low
    energy calculation directly with the fundamental quantum chromodynamics(QCD)
    makes effective theories still desirable.
As accepted widely, the relativistic nuclear theory can successfully describe the saturation
    at normal nuclear density and the spin-orbit splitting\refr{walecka1974,walecka1997}.
The further developments\refr{boguta1977} of $\sigma-\omega $ theory of quantum hadrodynamics model (QHD)
    make it possible to determine the model parameters
    analytically from a specified set of zero-temperature nuclear properties
    and allow us to study the hot nuclear properties and
     study variations of these results to nuclear
    compressibility or pairing correlation,
    and even the symmetry energy coefficient according to baryon density\refr{libao-an2000},
    which are not well known.
Furthermore, the in-medium hadronic property has attracted much attention with this kind of
    models and there are many existed works although the relation between
    QCD and QHD has not been well established\refr{shiomi1994,chen3}.
In recent years, with the refinement of nuclear theory study,
    the $\sigma -\omega $ theory has been placed in the context of effective theory
    and it is argued that  the {\em vacuum} physics has been explored in part by
    this kind of models\refr{brown1991,graciela1995}.
In physics, with the obvious
    non-vanishing fermion nucleon mass in relevant Lagrangian,
    the hidden chiral symmetry is explicitly broken.

The theoretical S-wave pairing correlation issue is a long-standing problem.
The fundamental $^1S_0$ pairing in infinite nuclear matter within
    the frame of relativistic nuclear field theory was first
    discussed by H.~Kucharek and P.~Ring\refr{ring1991},
    and it was found that the gaps are always
    larger for three times than the non-relativistic results\refr{gogny1980,matsuzaki1998}.
Especially, the very uncomfortable non-zero gaps of $^1S_0$ pairing correlation
    at zero baryon density obtained with frozen meson propagators in relativistic field theory,
    as recently pointed out by us\refr{chen4},
    remind us that the realistic nuclear ground state with MFT approach might not be EM empty. On the other hand,
    the well established low temperature superconductivity theory tells us that it would
    be very interesting to discuss the broken local EM symmetry effects on the properties of the
    nucleons system.
Although the in-medium nucleon-nucleon interaction potential induced by
    polarization can give a significantly improved description for EOS and superfluidity\refr{chen3,ji1988,chen4},
    the pairing difference of PP(proton-proton) from NP(neutron-proton) or NN(neutron-neutron) has been discarded.
Other approaches also recently found that polarization effects suppress the S-wave gaps
    by a factor of $3\sim 4$\refr{Schwenk2004}.
The numerical magnitude of $^1S_0$ gaps is not sensitive to a special parameters set
    and integral momentum cutoff when the polarization effect is taken into account\refr{chen4}.

The pure neutron matter cannot exist in nature, and the realistic
nuclear matter is subject to the long range EM
    interaction.
The changes of symmetry properties associated with possible phase transition
    realized on some conditions attract physicists very much.
In nuclear physics, charge symmetry breaking explored by the quite different
    empirical negative scattering lengths $a_{NN(P)}$
    and $a_{PP}$ is a fundamental fact\refr{duguet2003} and there are existed
    works to address its theoretical origin\refr{Gardner1995}.
    Coulomb correlation effects are a fundamental
    problem in nuclear physics and play an important role for the property of nuclear
    matter\refr{bulgac1999,Maruyama2004},
    which may lead to rich phase structures in the low temperature occasion.
For example, in Ref.\refr{gulminelli2003} the influence of the
non-saturating Coulomb interaction is
    recently incorporated in the multi-canonical formalism
    attempting to explain the reported experimental signatures of thermodynamic anomalies and the
    possible liquid-gas(LG) phase transitions of charged atomic clusters and nuclei\refr{siemens1983}.
One may naturally worry about the important role of the Coulomb
    repulsion force on the properties of charged system and the thermodynamics
    of charged/neutral nuclear matter to be reflected by relativistic nuclear theory and corresponding approaches.
Within the models based on $\sigma-\omega $
    field theory and usual adopted approaches such as MFT
    or relativistic Hartree approximation(RHA),
    one can {\em suppose} the similar interactions between PP  and NP or NN's,
    with  the {\em weak} EM interaction  being neglected compared to the
    residual strong interaction between nucleons.
Theoretically, the direct (Hartree) Coulomb contribution of charged protons to the EOS
    cannot be included due to the Furry theorem's limit.
Although the exchange (Fock) contribution can be included in principle from the point of view of field
    theory, the involved calculation and radioactive corrections caused by relevant infrared singularity of
    photon propagator still remain to be done even in the relatively simpler
    zero-temperature occasion in nuclear physics.
If one asks what the difference
    between PP and NP or NN pairing correlation is,
    the original version of QHD  with MFT or RHA approaches cannot tell us anything.
Although one can expect that the isospin breaking coupling terms such as $\rho NN$ etc.
    might reflect the Coulomb repulsion contribution on the thermodynamics of
    isospin asymmetric system to some extent, the pairing differences between PP and NN,
    NP exist even for symmetric nuclear matter
    incorporated with the quite different empirical negative scattering lengths $a_{NN(P)}$ and $a_{PP}$.
How to incorporate the important role of EM interaction with CSB on the thermodynamics
    of charged/neutral system or the property of nuclear ground
    state on a microscopic level (continuum field theory) remains an intriguing task even in an oversimplified way
    (MFT or RHA ) but with thermodynamics self-consistency.

In this Letter, we propose a systematic way to
    perform the link between the bulk and pairing correlation many-body properties of charged/neutral
    two-components nucleon systems through a relativistic nuclear field theory involving the interaction of
    Dirac nucleons with massive photons as well as the well known scalar/vector mesons.
Inspired by the continuum field theory of phase transition and
based on QHD-II, the constructed phenomenological Proca-like
Lagrangian through Anderson-Higgs mechanism is
\refr{peskin1995,Jenkins2003,walecka1974,walecka1997} \bea {\cal
L}=&&{\bar\psi }\[i \gamma _\mu \pa ^\mu -M-g_\sigma
\sigma-g_\omega \gamma _\mu \omega ^\mu \right.
\no\\
&&\left.
    -\012 g_\rho \gamma _\mu {\vec \tau}\cdot {\vec \rho} ^\mu
    -e \gamma _\mu \0{(1+\tau _3)}{2} A^\mu\]\psi
    \no\\
    &&
    +\012 \pa _\mu \sigma \pa ^\mu \sigma
    -\012 m_\sigma ^2 \sigma ^2
    -\014 H_{\mu\nu}H^{\mu\nu}
    \no\\
&&
    +\012 m_\omega ^2 \omega _\mu \omega ^\mu
    -\014 {\vec R}_{\mu\nu}\cdot {\vec R}^{\mu\nu}+\012
    m_\rho ^2 {\vec \rho}_\mu \cdot {\vec \rho} ^\mu
\no\\
&&
    -\014 F_{\mu\nu} F^{\mu\nu}
     +\012 m_\gamma^2 A_\mu A^\mu
    \no\\
    &&+\delta {\cal L} _{\small Higgs\& counterterm},
\eea
where $\sigma $,  $\omega ^\mu$, ${\vec \rho }^\mu$ and $A^\mu$ are the scalar-isoscalar,
    vector-isoscalar, vector-isovector meson fields, EM field with
    the field stresses
\bea
H_{\mu\nu}&&=\pa _\mu \om _\nu-\pa _\nu \om _\mu,\no\\
{\vec R}_{\mu\nu}&&=\pa _\mu {\vec \rho} _\nu-\pa _\nu {\vec \rho}
_\mu -g_\rho ({\vec \rho}_\mu \times {\vec \rho }_\nu ),
\no\\
F_{\mu\nu}&&=\pa _\mu A_\nu- \pa _\nu A_\mu \no \eea for $\om $,
${\vec \rho }$ and $A_\mu$'s, respectively. The $M$, $m_\sigma $,
$m_\om $, $m_\rho $ and $m_\gamma $
    are the nucleon, meson and photon masses, while $g_\sigma $, $g_\om $, $g_\rho $ and $e$
    are the coupling constants for corresponding Yukawa-like effective
    interaction, respectively.

Here the Lagrangian with CSB does not respect the local $U_{EM}(1)$ gauge symmetry which is
    broken by the ground state with non-zero
    local electric charge of protons
    (although the system can be globally neutralized by the surrounding
    such as electrons to maintain the stability for
    compact object through $\beta $-equilibrium).
Also, the quartic-cubic terms of $\sigma $ nonlinear-interaction potential
    $U(\sigma)=b\sigma^3+c\sigma^4$
    with the additional phenomenologically  determined parameters $b$ and $c$ have
    not been obviously preferred in order to discuss in a more general way
    although a specific assumption in $U(\sigma)$ can give a reasonable
    bulk compressibility for nuclear matter.

The mean field approximation can
    be used to discuss the thermodynamics of charged nuclear matter, from which the
    effective potential is derived in terms of finite temperature field
theory\refr{walecka1974,walecka1997,kapusta1989}
\bea
    \Omega/V
    =&&\012 m_\sigma^2\phi _0^2-\012 m_\omega^2 \omega_0^2-\012 m_\rho^2
    \rho_{03}^2-\012 m_\gamma^2 A_0 ^2
    \no\\
    &&-T \0{2}{(2\pi)^3} \sum _{i}\int d^3 {\bf k}
    \{\ln (1+e^{-\beta (E_i^*-\mu_i^*)})
    \no\\
    &&~~~~~~~~~~~~~~~~~~+\ln (1+e^{-\beta (E_i^*+\mu_i^*)})\},
\eea
where $i=P,N$ represents the index of proton (P) and neutron (N),
    respectively and $V$ is the volume of the system.
With the thermodynamics relation
\bea
    \epsilon=\01V \0{\pa (\beta \Omega)}{\pa \beta }+\sum _{i}\mu _i \rho
    _i,\no\eea
    one can obtain the energy density
\bea\label{energy}
    \epsilon =&&\0{m_\sigma^2}{2g_\sigma^2}(M-M^*)^2
    +\0{g_\omega^2}{2m_\omega^2} \rho _B^2
    \no\\
    &&+\0{g_\rho ^2}{8m_\rho^2} (\rho _P-\rho _N)^2
    +
    \0{e^2}{2m_\gamma ^2} \rho _P^2
    \no\\ &&
    +\0{2}{(2\pi )^3} \sum _{i}\int {d^3 {\bf k} }E_i^* \[n_i (\mu _i^*,T)+{\bar n_i}(\mu
    _i^*,T)\]\no\\
\eea
    and pressure $p=-\Omega/V$.
The baryon density is
\bea
    \rho _B&&=<{\bar \psi }\psi >_B=\sum _i \rho _i,\no\\
    \label{density}
    \rho _i&&=\0{2}{(2\pi )^3}\int d^3{\bf k} (n_i-{\bar n_i}).
\eea
In above expressions, $n_i (\mu _i^*,T)$, ${\bar n_i}(\mu _i^*,T)$ are the distribution functions for (anti-)particles
with $E_i^*=\sqrt{{\bf k}^2+M^{*2}}$.
The effective nucleon
    mass $M^*$,  chemical potentials $\mu_{P(N)} ^*$ are introduced by the tadpole
    diagrams of the sigma, omega, rho mesons and photon self-energies, respectively
    \bea
        M^*&&=M-\0{2}{(2\pi )^3}\0{g_\sigma^2}{m_\sigma^2}\sum _i\int d^3 {\bf k}\0{M^*}{E^*}(n_i +{\bar n}_i);\\
        \label{effchemical}
    \mu _P^*&&=\mu _P-\0{g_\omega^2}{m_\om ^2}\rho _B
        -\014 \0{g_\rho^2 }{m_\rho ^2}(\rho _P-\rho _N)-\0{e^2}{m_\gamma^2 }\rho _P,
        \no\\
    \mu_N^*&&=\mu _N-\0{g_\omega^2}{m_\omega^2}\rho _B+\014 \0{g_\rho^2 }{m_\rho ^2}(\rho _P-\rho _N),
    \eea
    where $\mu _{P(N)}$ is the proton (neutron) chemical potential.

\begin{figure}[ht]
        \centering
        \psfig{file=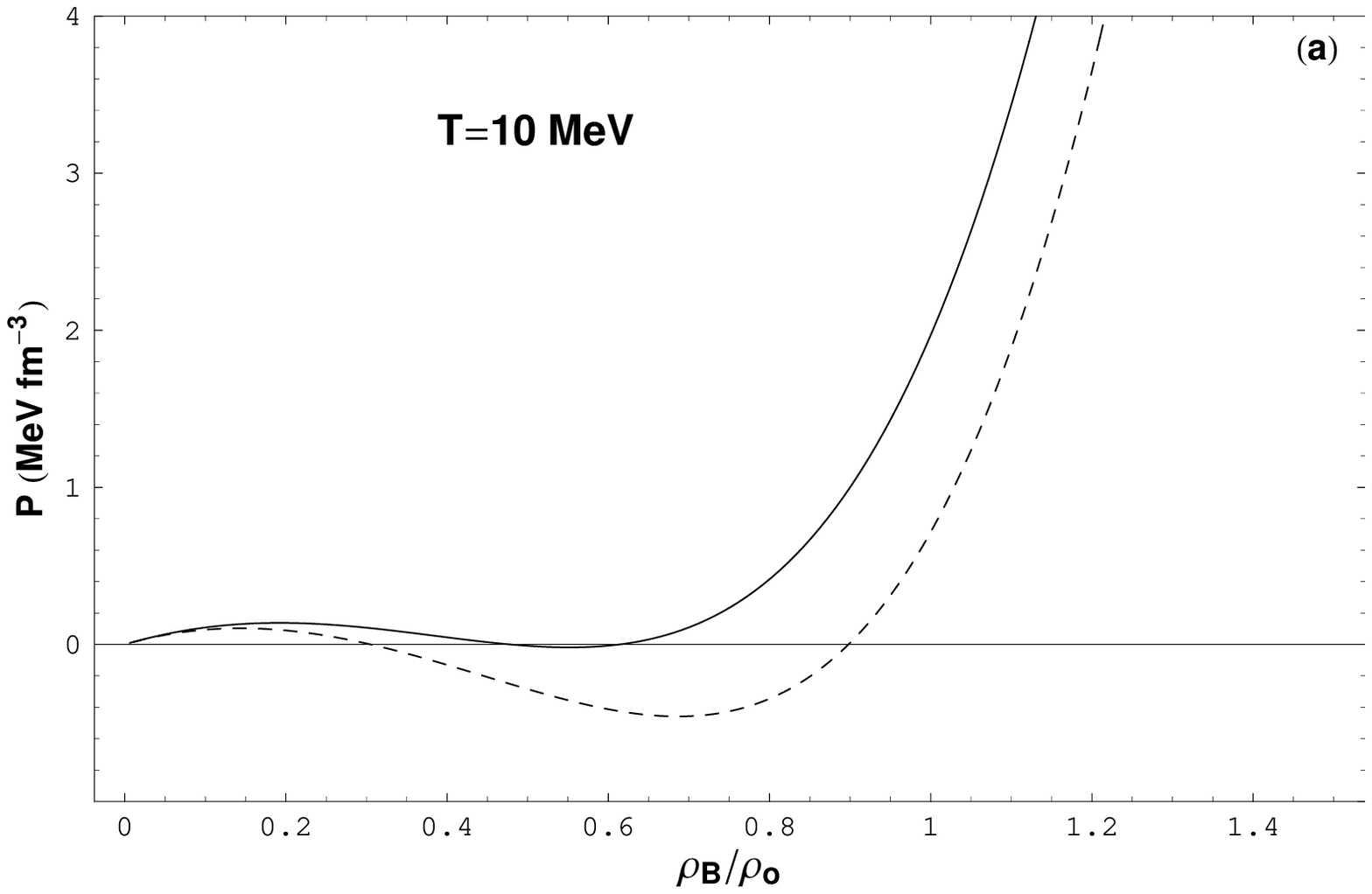,width=6.6cm,angle=-0}
        ~\\[.2cm]
    \psfig{file=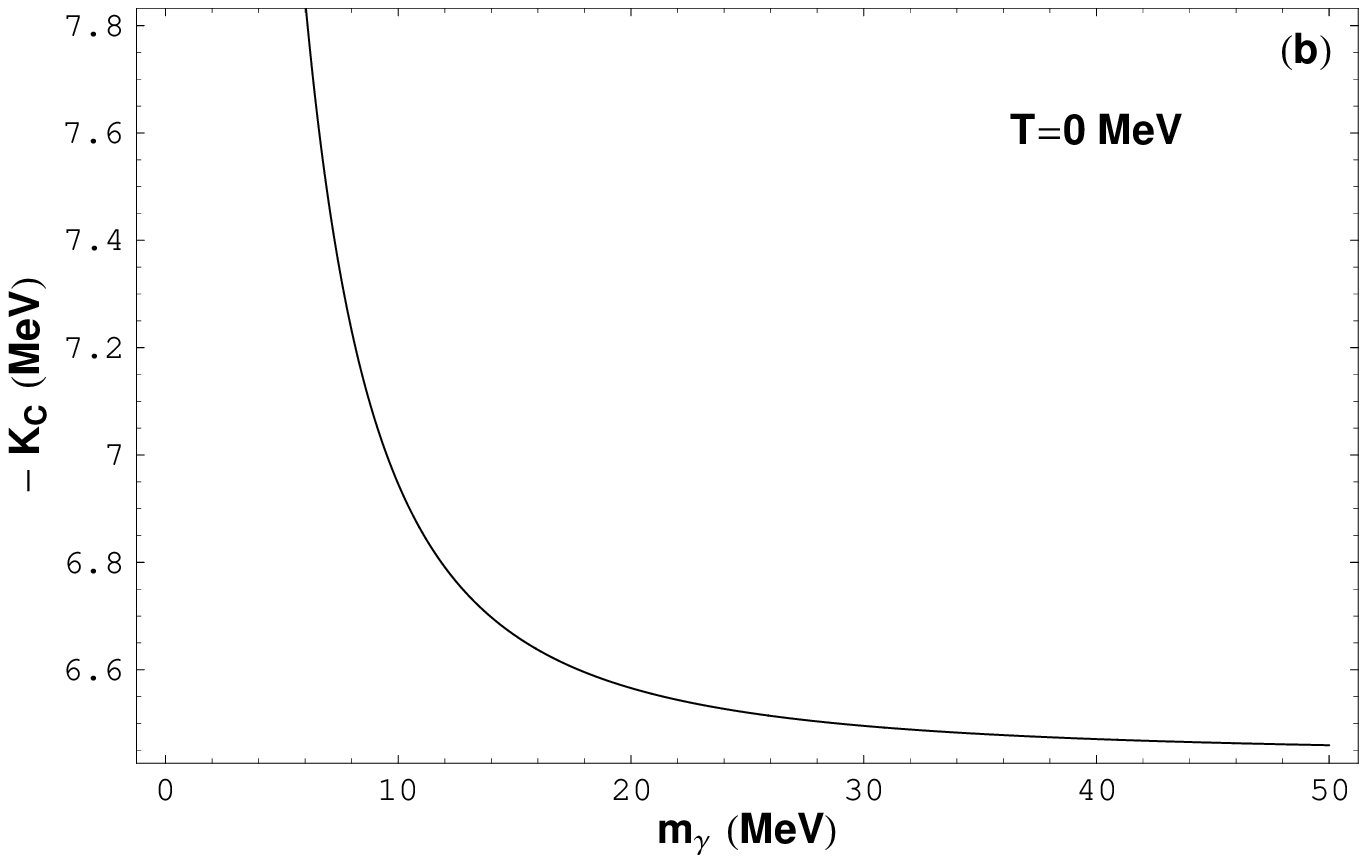,width=6.9cm,angle=-0}
        ~\\[.2cm]
    \caption{
        \small
For charged nuclear matter without considering the charge neutral condition
    with the  set (a) of Table.\ref{tab}:
(a)  Pressure versus rescaled density $\rho _B/\rho_0 $
    with (solid) and without (dashed) the Coulomb repulsion interaction.
(b) $K_C$ versus the order parameter $m_\gamma$ describing to what extent the EM symmetry is broken.}
\label{fig1}
\end{figure}

The photon mass $m_\gamma$ appears as a free parameter which
    is closely related to the Coulomb energy (reflecting the binding
    energy contributed by adding a proton {\em to} or removing a neutron {\em
    from} the system) and
    correspondingly to Coulomb compression modulus $K_C$.
It is worthy noting that the Coulomb energy can be discussed by the conventional many-body approaches such as the
    Thomas-Fermi theory with variational principle\refr{cheng1997}.
Within relativistic MFT,
    the $K_C$ has been analyzed in the literature such as in
    Ref.\refr{kuono1996} with the scaling model\refr{blaizot1980} phenomenologically.
The bulk compression modulus $K$ and $K_C$ are  defined by
\bea
    K=&&9 \rho _0^2 \0{\pa ^2 e_b}{\pa \rho _B^2 }|_{\rho_B =\rho
    _0},\no\\
    K_C =&&-\0{3 \alpha }{5 R_0 }(\0{9K'}{K}+8).
\eea
Here, $\rho _B $, $\rho _0$, $e_b$ are the baryon density, the normal baryon density
    and the binding energy per nucleon
    with \bea
    R_0=[3/(4 \pi \rho _0) ]^\013,~~~~
    K'=3\rho _0 ^3 \0{d^3 e_b }{d\rho _B^3 }|_{\rho_B =\rho
    _0}.\no\eea
The repulsive Coulomb force will modify the EOS significantly for the
    realistic charged system produced in heavy ion collisions.
Especially, it can make the critical temperature $T_c$ of the LG
    phase transition decreased to a smaller value.
With careful numerical
    study, it is found that the softness of bulk EOS (characterized by $K$) is
    not sensitive to the Coulomb interaction but the critical temperature $T_c$
    as well as $K_C$
    is very sensitive to this repulsive force.
The additional Coulomb energy
    term in pressure and energy density \eq{energy} contributes to removing the theoretical instability
    in the high baryon density region
    caused by a negative parameters set of $b$ and $c$ in the nonlinear
    self-interaction term $U(\sigma)$ (for obtaining a reasonable
    compressibility modulus of bulk EOS).
One can {\em estimate} the
    parameter $m_\gamma$ is about $20 -30 $ MeV for a reasonable critical
    temperature $T_c\sim 16 $ MeV accessible in heavy ion collisions.
In Fig.\ref{fig1}, we give the curves of pressure versus baryon
    density and the Coulomb compression modulus $K_C$ according to the
    interaction strength characterized by $m_\gamma$ with frozen parameters $g_\sigma $ and $g_\omega $ which
    are determined by fitting the binding energy $e^0_b=-15.75$ MeV and the bulk symmetry energy coefficient
    $a_{sym}=35$ MeV (for {\em symmetric} nuclear matter  at the
    empirical saturation density  $\rho
    _0=0.1484 fm ^{-3}$  with $T=0$)\refr{walecka1997}.
The qualitative Coulomb effect on the deformation of phase space distribution functions resulting
    from \eq{density} and \eq{effchemical}
    can be reflected by the proton fraction ratio: $Y_P=\rho _P/\rho _B$.
It is found that this ratio changes significantly according to temperature $T$ and total baryon density $\rho _B$.

Therefore the electric repulsive force plays an isospin violating role for the many-body property.
Indeed, there is some kind competition between the $\rho $ and photon's isospin breaking effect on the
    phase space distribution function {\em deformation}.
The Coulomb force makes the
    proton fraction decreased while the $\rho $ meson plays a weak inverse role.
Furthermore, one can readily derive the symmetry energy coefficient formula at $T=0$
\bea\label{sym}
a_{sym}&&=\0{1}{2}\0{\pa ^2 (\epsilon /\rho )}{\pa t^2}{\large |}_{t=0}\no\\
&&=\0{k_f^3}{12 \pi ^2} (\0{g_\rho ^2}{m_\rho ^2}+\0{e^2}{m_\gamma
^2})+\0{k_f^2}{6\sqrt{k_f^2+{M^{*}}^2}},\\
t&&=\0{\rho _N-\rho _P}{\rho _B}.\no
\eea
In fact, if without taking into account the repulsive Coulomb contribution, one must introduce a
    {\em very large} coupling constant $g_\rho$ to approach the empirical symmetric
    coefficient $a_{sym}$ which is very far from the empirical coupling
    constant $g_{\rho NN}$ extracted experimentally.
From \eq{sym}, if taking $g_\rho =2.63$\refr{shiomi1994} and $a_{sym}=35$ MeV, the
    free parameter $m_\gamma$ can be fixed accordingly.
With close study, the numerical magnitude of
    $m_\gamma $ is more sensitive to $M^*$ (and hence the softness of bulk EOS )
    than  to $g_\rho$.
The relevant parameters are listed in Table.\ref{tab} corresponding to the L2 set of Ref.\refr{walecka1997} except
    of taking $g_\rho =2.63$.
Further exact fitting to finite nuclei data such as charge density radii distribution
    etc.\refr{walecka1997,horowitz2001} and addressing the spin-orbit splitting issue with
    the mirror-symmetry topic can contribute to giving a solid limit for fixing $m_\gamma$ and $g_\rho$.
The large tensor and spin-orbit forces are also crucial for understanding finite nuclei
    and neutron star structure which can be explored through the study for mirror-nuclei.
Let us mention that the hitherto overlooked but important EM interaction role on the spin-orbit
    splitting for some mirror-nuclei is recently found in Ref.\refr{Ekman2004}.

\begin{figure}[ht]
        \centering
        \psfig{file=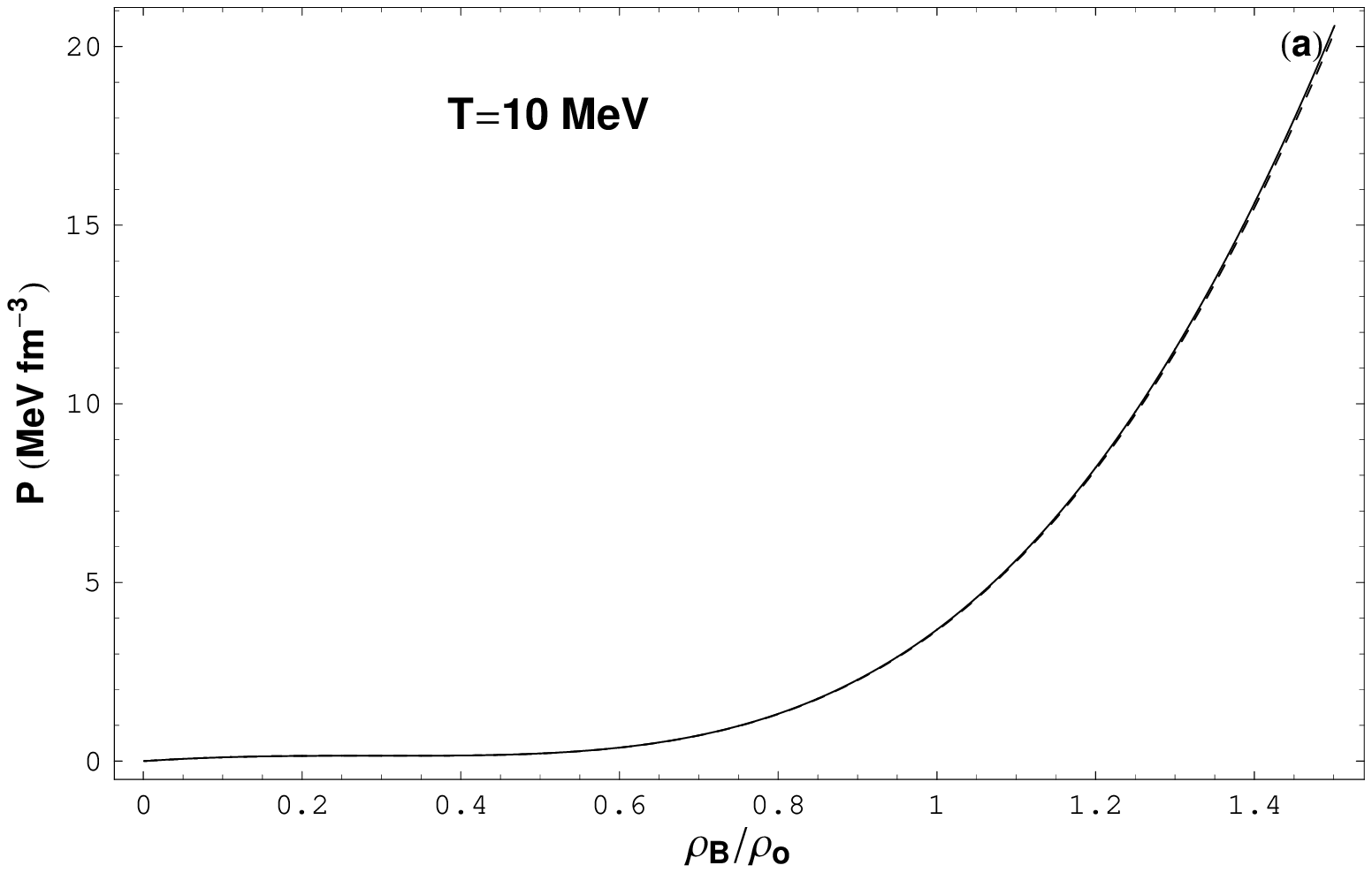,width=6.6cm,angle=-0}
    \psfig{file=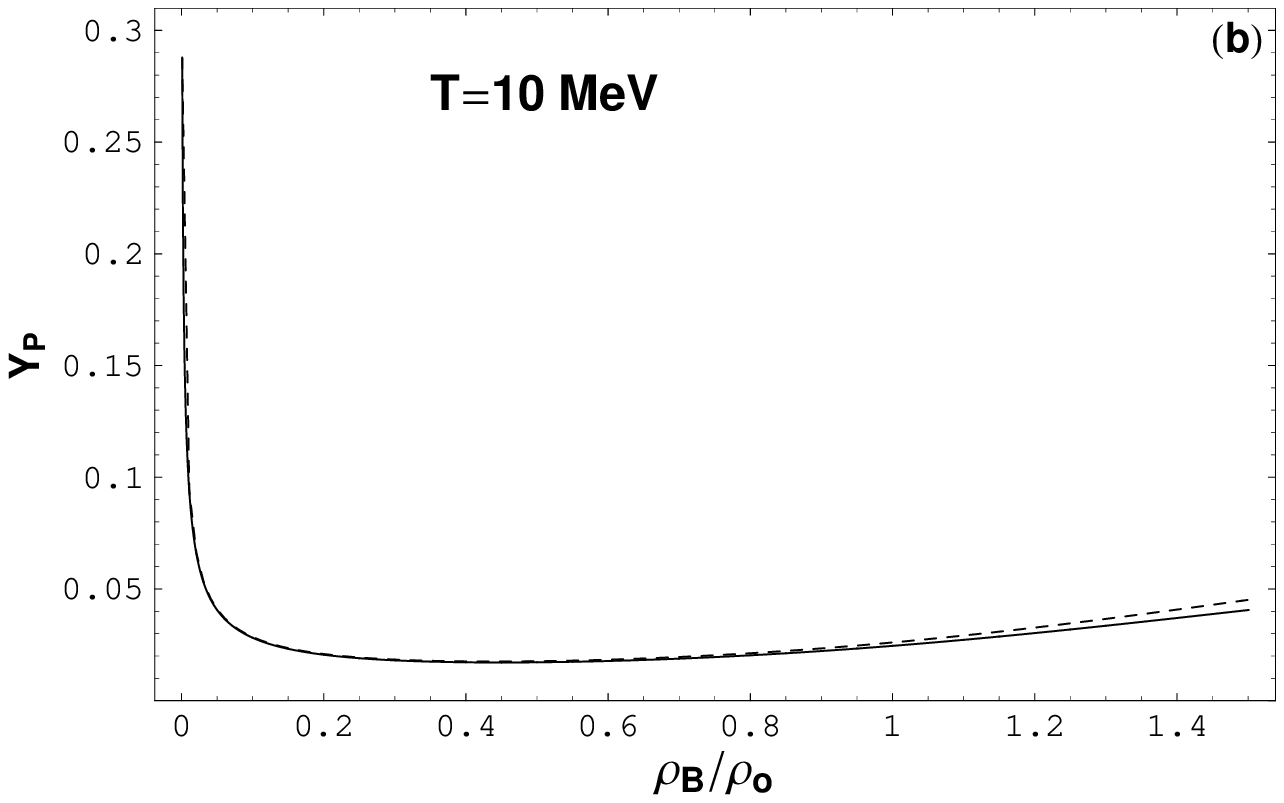,width=6.9cm,angle=-0}
        ~\\[.2cm]
    \caption{
        \small
For electric neutral nuclear matter stabilized through $\beta$-equilibrium:
    (a)  Pressure versus rescaled density.
    (b) Proton fraction ratio $Y_P$. Line-styles are similar to Fig.\ref{fig1}.}
    \label{fig2}
\end{figure}

To study its effect on the electric neutral nuclear matter such as compact proto-neutron star
    would also be interesting.
For the simplest $NP+e+\nu _e$ system stabilized through $\beta $-equilibrium, charge
    neutrality condition makes the proton fraction ratio very small.
It is found that the Coulomb force does not modify this picture as indicated by Fig.\ref{fig2}.
This is consistent with above result that the Coulomb interaction doesn't change the softness of
    bulk EOS significantly.

Correlations not only do manifest themselves in the bulk properties but also
    modify the quasi-particle properties of nucleons in a substantial way.
Conceptually,
    the PP and NN(P) pairing correlations should be quite different from each other.
The former has additional superconductivity contribution due to the electric charge of protons in
    addition to the attractive residual strong interaction compared with the
    scenario of NN correlation.
For the fundamental $^1S_0$ pairing,
    the energy gap equation of nucleon-nucleon pairing in the frame of relativistic nuclear
    theory can be reduced to\refr{ring1991,matsuzaki1998,chen4}
\bea\label{gap}
    \Delta (p)=-\0{1}{8\pi ^2 }\int {\bar v}_{pp} (p,k) \0{\Delta (k)}
    {\sqrt{\varepsilon (k)^2 +\Delta ^2(k)}}k^2 dk,\no\\
\eea
    and the coupled effective mass gap equation has been neglected here for brevity.

\begin{figure}[ht]
        \centering
    \psfig{file=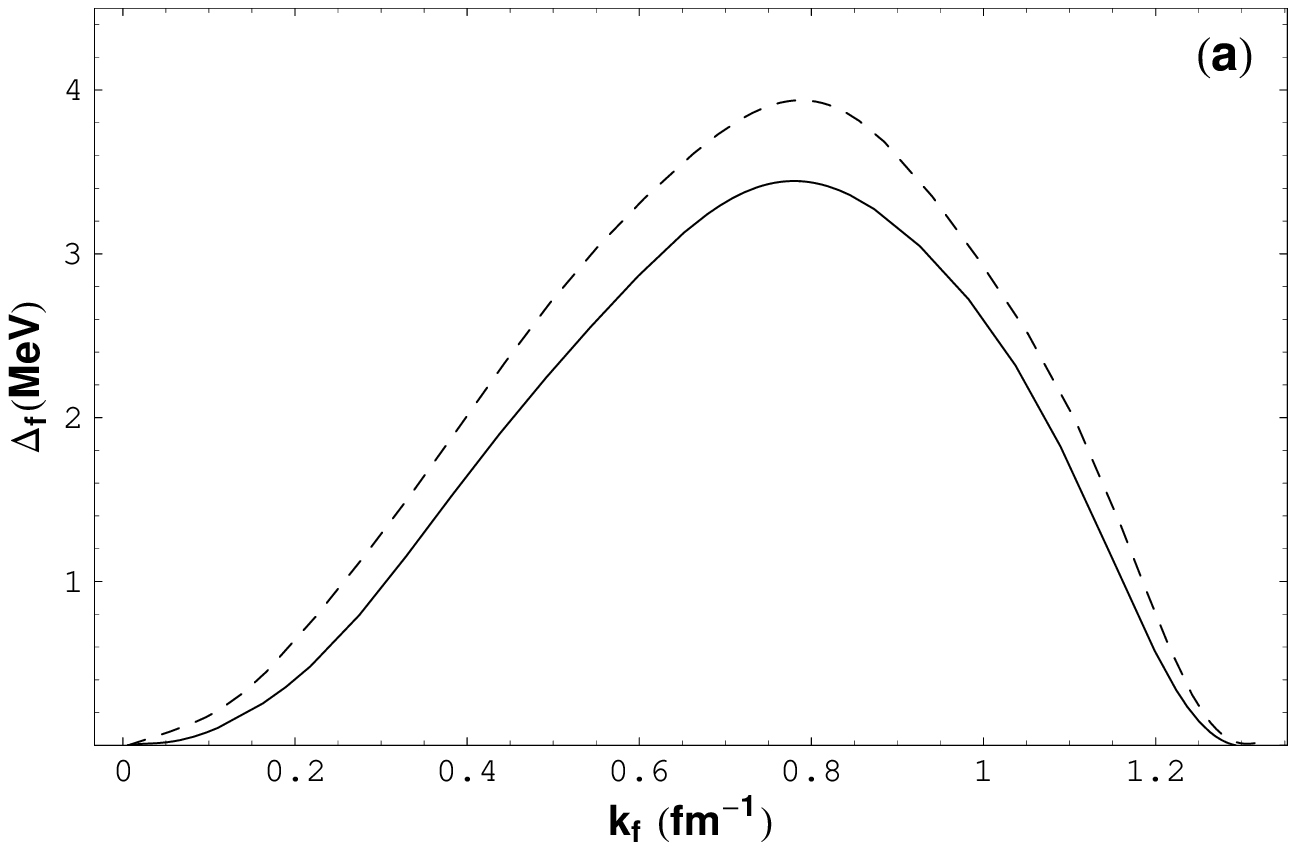,width=6.9cm,angle=-0}
        \psfig{file=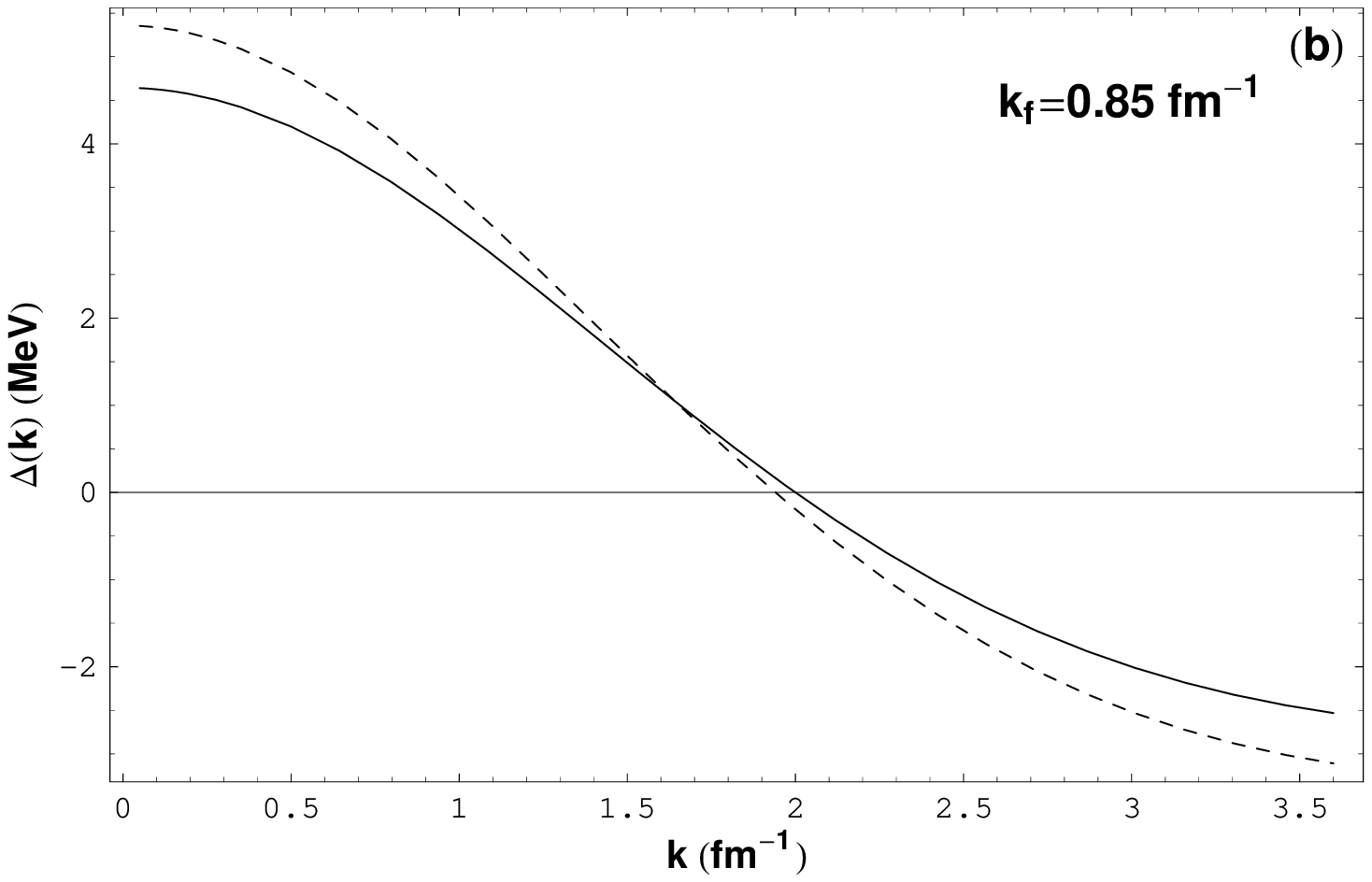,width=6.6cm,angle=-0}
        \caption{
        \small (a) Pairing gap $\Delta _f$ at the Fermi surface versus Fermi momentum $k_f$. (b)
        Gap function $\Delta (k)$ versus momentum $k$ for fixed Fermi momentum. The solid line corresponds to the result of
        proton-proton pairing correlation and dashed line to that of neutron-neutron.
        }\label{fig3}
\end{figure}

In \eq{gap},
    the assymetrized matrix elements ${\bar v_{pp}} (p,k) $
    is obtained through the integration of ${\bar v}(\vp,\vk)$
    over the angle $\theta $ between the three-momentums $\vp$ and $\vk$ with
    ${\bar v}(\vp,\vk)$ being the particle-particle interaction potential
\bea
    \bar{v }(\vp,\vk)
        =&&\mp \0{M^{*^2}}{2 E^*(k) E^*(p)}
        \no\\
        &&\0{\Tr [\Lambda _+ (\vk )
    \Gamma
        \Lambda_+ (\vp )\gamma ^0
        {\cal T}^+ \Gamma  ^+{\cal T}\gamma^0]}{(\vk -\vp )^2 +m_D^2}\no,
\eea
where $\Lambda _+ (\vk )=\0{\fslash{k}+M^*}{2M^*}$ is the
    projection operator of the positive energy solution and ${\cal T}=i \gamma^1 \gamma ^3$ is the time reversal operator.
The $\Gamma $ is the corresponding interaction vertex of $\sigma
    $/$\om, \rho (\gamma )$ with nucleons while $m_\gamma $ is the {\em photon
    mass}.
This static electric contribution to the gap has been indicated in
    Fig.\ref{fig3} as a curve of gap versus density with a integral momentum
    cutoff $\Lambda _k=3.6 fm^{-1}$ and the set (a) in Table.\ref{tab} to numerically solve the integral gap equation.
The difference between
    PP and NN pairing correlations reflects that virtual photons
    proceeding in the space-like momentum transfer regime carry a
    unique information on the EM properties of nucleon interaction responsible for
    the nucleon structure.
As indicated by Fig.\ref{fig3}, the long range but screened Coulomb interaction affects the
    correlation function of proton-proton pairing significantly esp. in
the low momentum regime.

The physical reason for the parametric description of the EM interaction in this approach
    is that at first one can note the existence of locally charged system/cluster, i.e.,
    the electromagnetic field condensation $\sim <{\bar \psi _P} \gamma
    ^\mu \psi _P>$
    (corresponding to the spontaneously breaking of local gauge symmetry while
    the gauge field obtains mass) in the low energy scale although the
    stable system should be globally neutral with surrounding such as electrons through
    $\beta$-equilibrium.
This is very much similar to the chiral condensation $\sim <{\bar \psi } \psi >$ at
    low energy scale. Second, from the point of view of continuum field theory with
    symmetry changes, the
    physics background of well known low temperature LG phase transition still
    remains to be explored.
Especially, how to reflect the CSB characteristic in relevant effective theory
    and approaches remains to be performed.
Third, in the multi-components Fermi/Bose systems the CSB would lead to more rich phenomena, e.g., compared with the
    metal electric superconductivity occasion(the ions fixed as lattice).

From the point of view of Maxwell QED, because photons are massless, photon-mediated interactions are long range in contrast with a point-like meson-nucleon interaction
    in the existed QHD-like Lagrangian.
The long range nature of photon-exchange manifests itself in the infrared singular
    behavior of the photon propagator.
This characteristic enhances the contribution of very soft,
    collinear photons to the correlation energy for the EOS or the pairing
    problem by noting that this divergence should be avoided by the resummation
    approach as done in QCD or QED.
Essentially, different from the QCD occasion(with the magnetic
    mass cutoff due to the {\em non-Abelian} self-interaction of gluons)\refr{Rischke2000},
    there is no magnetic screening in QED, which makes it very
    involved to discuss the superconductive behaviors in strong magnetic field
    occasion such as in compact star environment with conventional QHD-like Lagrangian.
This approach makes it possible to further study Meissner as well as Debye screening
    effects in such as astrophysics\refr{Buckley2004}.
The premise of this approach as a nonlocal effective theory nearby a phase transition with CSB
    through one-meson(photon)-exchange picture would be very powerful in addressing
    nuclear matter (either symmetric or asymmetric) many-body properties and even
those of finite nuclei.

In summary,
    the long range non-saturating Coulomb interaction plays an important role
    in the property of charged/neutral nuclear matter,
    which can be incorporated simultaneously with the residual
    strong interaction within MFT of relativistic nuclear theory through
an effective Proca-like Lagrangian.
The deformation of phase-space distribution function of nucleons attributed
    to the static electric interaction can manifest itself on the
    thermodynamics or the property of ground
    state of charged nuclear matter and the quasi-particle spectrum while
    the photon mass $m_\gamma$ controls the strength.
Especially, the repulsive Coulomb force makes
    the critical temperature lower than the existed
    theoretical anticipation and contributes to interpreting the accessible
    experimental results.
Furthermore,
    the breached PP and NN pairing
    correlation strengths open a new window for the study of nuclear matter
EM property/nuclei
    structure and would lead to rich physical phenomena.
From the view of point of many-body physics,
    the low-temperature LG phase transition found in heavy ion collisions and
    the different correlation strengths for PP and NN(P) bound-state can be seen as the
    fingerprint of broken EM symmetry within MFT to some extent.
Our discussion based on assuming the spontaneously broken
    EM gauge symmetry highlights that the $U(1)$ electric charge symmetry
    violating effects should be taken into account simultaneously with the
    $SU(2)$ isospin breaking effects played by such as $\rho NN$ coupling.
The  weak coupling interaction is mixed with other stronger ones and plays
    an important role for the many-body effects.
Especially, the
    overlooked EM interaction contribution on the many-body properties such as thermodynamics,
    binding, pairing mechanism etc. of nucleons in nuclear matter
    should be carefully considered from the point of view of continuum field theory.
{\bf Acknowledgments}: Ji-sheng Chen acknowledges the
    beneficial discussions with Prof. Hong-an Peng(Peking University).
This work was supported by NSFC  under grant Nos. 10175026,
90303007.
\begin{table}
\caption{
The parameters are with $M=939$, $m_\rho =770$, $m_\om =783$,
        $m_\si =520$ MeV(s) and $m_\gamma $ is in (MeV). $C_i^2=g_i^2 M^2/m_i^2$.}
\begin{ruledtabular}
   \begin{tabular}{ccccccccc}
 Set& $g_\sigma^2 $ & $g_\om^2$  & $g_\rho ^2(C_\rho ^2)$&$m_\gamma(C_\gamma ^2)$&
$\0{M^*}{M}|_{\rho _0}$&
\\
\colrule
        MFT\\
 a&91.64&191.05 &6.91 (10.28)&30.44 (87.27)&0.540&\\
        b&     &       &0           &28.79 (97.55)&&  \\
            c&    &       & 65.58 (97.55)& $\infty$ or e=0 (0)&& \\
\colrule
        RHA \\
d&69.98&102.76 &6.91
(10.28)&26.636 (113.96)&0.731&\\
        e&     &       &0           &25.51 (124.24)&&  \\
        f&     &       & 83.54 (124.24)           & $\infty$ or e=0 (0)&&  \\
\end{tabular}
\end{ruledtabular}
\label{tab}
\end{table}


\end{document}